\documentstyle[aps,prl]{revtex}

\def\kap{{\textstyle{1\over{\kappa^2}}}}
\def\half{{\textstyle{1\over2}}}
\def\thre{{\textstyle{1\over3}}}
\def\twtr{{\textstyle{2\over3}}}

\def\quart{{\textstyle{1\over4}}}
\def\eigt{{\textstyle{1\over{2^3}}}}
\def\ninsix{{\textstyle{{{1}\over{3\cdot 2^6}}}}}

\begin{document}

\draft
\title{
Spontaneous Breaking of Diffeomorphism Invariance in Matrix Theory}

\author{Shyamoli Chaudhuri$^{\dagger}$}

\address{Laboratory of Nuclear Studies, Cornell University, Ithaca 
NY 14853 \\ 
Institute for Theoretical Physics, University of California at Santa Barbara,
Santa Barbara, CA 93106}

\date{\today}

\maketitle

\begin{abstract}
We present a matrix action based on the unitary 
group $U(N)$ whose large $N$ ground states are conjectured to be in precise 
correspondence with the weak-strong dual effective field theory limits of M theory preserving
sixteen supersymmetries. We identify a finite $N$ matrix algebra that corresponds to 
the spacetime and internal symmetry algebra of the Lorentz invariant field theories 
obtained in the different large $N$ limits. The manifest diffeomorphism invariance of 
matrix theory is spontaneously broken upon specification of the large $N$ ground state. 
We verify that there exist planar limits which yield the low energy spacetime 
effective actions of all five supersymmetric string theories in nine spacetime 
dimensions and with sixteen supercharges. 
\end{abstract}
\pacs{}

\vskip 0.1in
The consistent perturbative quantization of supersymmetric Yang-Mills theory with 
chiral matter coupled to supergravity given by either the Type I or Heterotic String 
Theory is a remarkable achievement \cite{polbook}. However, in the absence 
of a nonperturbative framework there remains open the question of an underlying
fundamental principle \cite{polbook,wit,dbrane}. In previous matrix model proposals 
\cite{bfss,ikkt} that attempt to address this issue, an implicit assumption made
was that the crucial long distance gravitational interaction would arise from 
off-diagonal $O(1/N)$ quantum corrections to the planar matrix action. The 
success of either matrix model proposal in meeting this consistency check remains 
in doubt \cite{dine}. 

\vskip 0.1in
Equally serious is the absence of a demonstration of spacetime Lorentz invariance in 
the large $N$ limit of the Banks-Fischler-Shenker-Susskind matrix model \cite{bfss}. 
In the Ishibashi-Kawai-Kitazawa-Tsuchiya matrix model \cite{ikkt} on the other hand, 
the arguments for making contact 
with the $\cal N$$=$$2$ supersymmetries of the ten-dimensional IIB string and its 
Dstring ground states remain shaky \cite{ikkt}. A clear distinction of Ramond 
sector solitons from the Neveu-Schwarz sector solitons of the supersymmetric gauge 
theory is lacking. Nevertheless, the intriguing evidence of a dynamical mechanism in the 
matrix model which preferentially selects ground states with fewer than ten noncompact 
spacetime dimensions deserves further study \cite{nishimura}.

\vskip 0.1in
In this paper, we resolve these issues directly at the classical level by proposing 
a new action for matrix theory which preserves sixteen supercharges, and is based on both adjoint 
and vector multiplets of the large $N$ group $U(N)$. The planar limits of the matrix action 
correspond to classical ground states of String/M theory. Quantum $1/N$ corrections to the 
planar action are to be directly identified with quantum and nonperturbative corrections 
to classical string theory, both being absent in the special case of string theories which
preserve sixteen supersymmetries. The planar limits of our matrix theory are manifestly 
Lorentz invariant. They can exhibit local supersymmetry, spacetime chirality, nonabelian 
gauge symmetry, and any additional higher rank tensor gauge symmetries characteristic of a 
supersymmetric effective field theory action. We will restrict ourselves in this paper to 
recovering all five string theories in nine noncompact dimensions as different large $N$ 
limits of matrix theory: IB and heterotic $O(32)$, heterotic $E_8$$\times$$E_8$, 
and the massive IIA and IIB strings \cite{wit,polbook}, each preserving sixteen 
supersymmetries, and having a manifestly Lorentz covariant Lagrangian description. 

\vskip 0.1in
Our proposed action for matrix theory takes the form:
\begin{eqnarray}
{\cal S} &&=
\half \kap \left ( {\bar \psi}_{a} \Gamma^{abc} D_{b} \psi_{c}
- 4 {\bar{\lambda}} \Gamma^{ab} D_{a} \psi_{b}
- 4 {\bar{\lambda}} \Gamma^{a} D_{a} \lambda \right )
\cr
\quad &&
+ \half g^2 e^{\Phi} {\bar \chi}^i \Gamma^{a} D_{a} \chi^i
+ \half g^2 e^{\Phi } {F}^{ab} F_{ab}
+ \half \kap ( {\cal R}
\cr
\quad && \quad
- 4 {\Omega}^{a} \Phi  \Omega_{a} \Phi
+ 3 e^{2 \Phi}  {H}^{abc} {H}_{abc} )
+   {\cal S}_{\rm 2}  + {\cal S}_{\rm 4}  .
\label{eq:hmat}
\end{eqnarray}
$S_2$ and $S_4$ will denote, respectively, two-fermi and four-fermi matrix 
operators, identified below by requiring invariance of the classical matrix
action under the finite $N$ symmetry algebra $\cal G$. $\cal G$ plays the 
same role at finite $N$ as the combined group of spacetime and internal symmetries 
found in 
the large $N$ limit--- a Lorentz invariant, supersymmetric, nonabelian 
gauge-gravity field theory with up to sixteen conserved supercharges. 
It should be apparent from Eq.\ (\ref{eq:hmat}) that $\cal S$ has many 
inequivalent large $N$ ground states: in addition to taking $N$ large, we 
can hold fixed certain dimensionless combinations of background fields and 
couplings. The resulting multitude of large $N$ ground states is conjectured 
to be in precise accord with the expected multitude of weak-strong dual 
effective field theory limits of M theory \cite{wit}. This is a natural 
generalization of the double scaling limit defined for the two-parameter 
one-matrix model \cite{mat}. 

\vskip 0.1in
The fundamental 
variables in our nonperturbative framework are, bosonic and fermionic, hermitian 
matrices living, respectively, in the $N$$\times$$N$ adjoint, and $N$-dimensional 
vector representations of the group $U(N)$. 
The introduction of Grassmann-valued matrices transforming in 
the $N$ and $\bar N$ fundamental representations of the $SU(N)$ subgroup allows 
us to incorporate a crucial feature absent in previous
matrix model proposals \cite{bfss,ikkt}--- {\em chirality}. The 
large $N$ limit of our action straightforwardly 
gives rise to sixteen component spinors which simultaneously satisfy both 
the Majorana and Weyl conditions of $d$$=$$10$ $\cal N$$=$$1$ SYM-supergravity 
\cite{fs,br}. A large $N$ ground state of our matrix theory corresponds 
to diagonalized configurations for all $SU(N)$ matrices. The dimensionful couplings 
and fields of the large $N$ limit are scaled with respect to the 11d Planck 
length, or inverse closed string tension, which we will hold fixed: 
$ l_{11}$$=$$ N \epsilon $$=$$N e^{ \twtr \Phi_0} \alpha^{\prime 1/2}$, where 
$\Phi_0$ is the vev of the dilaton scalar in the specified ground state of matrix
theory, and $g_c$$=$$e^{\Phi_0}$ is the closed string coupling. 
Equivalently, we can define the size of the eleventh dimension, 
$R_{11}$$=$$e^{\Phi_0} \alpha^{\prime 1/2}$. Unlike the matrix models in 
\cite{bfss,ikkt}, the appearance 
of a continuum spacetime manifold in the large $N$ limit will be {\em implicit}: 
a noncompact dimension might be evidenced, for example, by observation of the 
onset of a continuum spectrum of eigenvalues for the scalar Laplacian, 
$\Delta$$\equiv$$\Omega^a \Omega_a$, where $\Delta$ is, more generally, at finite
$N$, a self-adjoint linear operator acting on bosonic hermitian matrices defined
on the field of real (complex) numbers.

\vskip 0.1in
Notice that spacetime diffeomorphism invariance follows automatically in 
any nonperturbative matrix formulation for a quantum theory of gravity based on 
{\em zero-dimensional} matrices. It is only in the specification of the 
ground state, namely, the large $N$ limit about which we expand when computing the 
$O(1/N)$ quantum corrections arising from the matrix path integral, that diffeomorphism
invariance is {\em spontaneously} broken: the norm in the eigenspace of any 
self-adjoint operator is specified with respect to a fixed background metric. Thus,
the perturbative string theories and their low energy effective field theory 
limits describe long-distance fluctuations of matter-energy in a spacetime continuum 
with fixed background metric, and in a ground state of matrix theory which spontaneously 
breaks diffeomorphism invariance. Notice that we have preserved the spirit of 
Einstein's classical theory of general relativity \cite{einstein}: 
spacetime geometry is replaced by an 
equivalent matter-energy distribution, together with the classical field equations.
But given both the matter-energy distribution and a specific solution to the field
equations enables recovery, at least in principle, of a corresponding 
background metric. 

\vskip 0.1in
The reader might be tempted, at this juncture, to interpret our action ${\cal S}$ 
as a classical Hamiltonian. The natural extension to a quantum theory 
corresponds to quantum mechanics in the Heisenberg representation with an
underlying operator algebra contained within our algebra $\cal G$. 
The matrix realization of 
self-adjoint linear operators given by $\cal S$ is perhaps best understood 
within the representation-independent framework of Dirac's quantum mechanics, 
or of noncommutative geometry \cite{dirac}.
\footnote{These remarks were 
inspired by some beautiful ongoing lectures by I.\ M.\ Singer.} 

\vskip 0.1in
To illustrate the basic methodology we will follow,
consider the hermitian one-matrix model \cite{bipz}:
\begin{equation}
{\cal S_B} = {\rm Tr} ~ 
\Phi \Omega^a \Omega_a \Phi   , \quad \Omega_a \Phi_n = \lambda^{1/2}_{na} \Phi_n   
\label{eq:scalar}
\end{equation} 
with $ n $$=$$ 1,\cdots , N $. 
We are assuming here that $\Delta$$=$$\Omega^a \Omega_a$ is a 
self-adjoint linear operator $\Delta$, with orthonormal generalized eigenstates, 
$\{ \Phi_n \}$, and eigenvalues, $\{ \lambda_n \}$. 
In the large $N$ limit, we make the usual 
extension to a Hilbert space with continuous eigenfunctions 
$\{ \phi(\lambda ; {\bf x}) \}$. Note that the dependence on the underlying 
target space, ${\bf x}$, need not be explicit: we can work in a generalized momentum 
basis, or with suitable phase space variables. In general, the transition from 
finite $N$ to infinite $N$ is not unique. Given 
the nature of the eigenvalue spectrum of $\Delta$ in the large $N$ limit, 
appropriate low and high momentum regulators may be needed in order to make the 
integral well-defined. These basic guidelines hold for all of the 
kinetic terms in our matrix action.

\vskip 0.1in
More importantly, unlike the path integral of the one-matrix model where
both the action and measure readily localize on the 
space of eigenmodes \cite{bipz,mat}, $\cal S$
has nontrivial terms coupling two or more matrices. Recall that no 
symmetry other than $SU(N)$ invariance was exploited in the
analytic solution of the one-matrix path integral
\cite{bipz}. What happens if the matrix action is characterized by 
additional symmetry?  This requires careful analysis of the invariant 
measure for the semi-direct product group $\cal G$$\times$$SU(N)$. 
It may be helpful to keep in mind that the analytic solution in
\cite{bipz} could have been obtained as follows: carry out the matrix 
path integration in closed form {\em prior} to performing the 
large $N$ expansion, and subsequent term-by-term diagonalization. 
The Feynman path integral denotes the 
sum over a complete set of matrix-valued 
orthonormal eigenstates for some given self-adjoint linear operator. 
This result is basis independent. In particular, we need not work with
the usual position-space eigenfunctions of field theory Lagrangians, 
choosing instead to work in a generalized momentum basis.
Thus, for example, we switch off any interactions to obtain
${\rm tr}_{\{\Phi_n \}} ~ e^{-{\Phi} \Delta \Phi} 
 $$ \equiv $$[\det \Delta ]^{-1/2} $.
Expanding in a basis of free eigenmodes, we can now perform a 
Feynman diagram expansion invoking perturbation theory in a suitable 
dimensionless parameter or coupling present in the interaction. 
The large $N$ limit--- or some multiple-scaled variation on double
scaling \cite{mat}, and the computation 
of $O(1/N)$ corrections, are the final steps of this computation 
giving concrete meaning to the matrix quantum effective action. 

\vskip 0.15in
The inner product in each eigenspace will be 
defined by ordinary matrix multiplication. We distinguish left- and right- 
multiplication within any matrix bilinear. The ordering ambiguity is fixed
by recognizing the $SL(2,C)$ decomposition of each operator:
objects in the $(n,0)$, or $(0,{\bar {n}})$, representation are left-multipliers, while
the $(0,n)$, or $({\bar{n}} , 0)$ representations are right-multipliers. 
Thus, anti-spinors precede spinors and contravariant tensors 
precede covariant tensors in any bilinear. 
For example, $\eta^{ab} $$\equiv$$ E^{a \mu} E^{b}_{\mu}$, 
${A}^{\mu} A_{\mu} $$=$$ {A}^a E^{\mu}_a E_{b \mu} A^b $$=$$ {A}^a A_a $,
with spinors defined by: $ {\bar{\psi}}_{\mu} $$=$${\bar{\psi}}_{a}  
E^{a}_{\mu}$, $ \psi_{\mu} = E^a_{\mu} \psi_{a}$.
Note that spinor and anti-spinor
denote, respectively, fermionic Grassmann-valued objects 
in the $N$ and ${\bar{N}}$ 
representations of $SU(N)$. 

\vskip 0.1in
The fermionic operator $(\Gamma^a D_a)_{\alpha \beta}$ can be expressed
as a sum over the independent Lorentz structures in theories with sixteen 
supercharges:
\begin{eqnarray}
\Omega_a (\Gamma^a)_{\alpha\beta}
+&& \Omega_{ab} ( \Gamma^a \Gamma^{b})_{\alpha\beta}
+ \Omega_{abc} (\Gamma^a \Gamma^{bc})_{\alpha\beta}
\cr
\quad +&& \Omega_{abcd} (\Gamma^a \Gamma^{bcd})_{\alpha\beta}
+ \Omega_{abcde} (\Gamma^a \Gamma^{bcde})_{\alpha\beta}
\label{eq:covmateq}
\end{eqnarray}
$\Gamma^a D_a$ evolves into the super-covariant derivative in
the large $N$ limit, coupling to
both spin connection and nonabelian vector potential. 
It could also couple to additional antisymmetric $p$-form
gauge potentials of the nonperturbative string or field theory
ground state \cite{dbrane,polbook}. Such couplings
represent Dbrane vacuum charges \cite{dbrane}, necessitated 
by the requirement that the family of matrix theory ground 
states is also closed under weak-strong coupling duality 
transformations \cite{wit,polbook}.
For example, in the generic IB ground state, $\Gamma^a D_a$
would evolve to:
\begin{eqnarray}
( {\bf 1} +&& C_{(0)}(x) ) \partial_{a}  +
 A_{a}^{j} (x) \tau^{j}  ) \Gamma^a
+ ( A_{ab} (x) 
\cr
\quad +&& C_{ab} (x) ) \Gamma^a \Gamma^{b}
 +  \omega_{abc} (x) \Gamma^a \Gamma^{bc}
+  C_{abcd} (x) \Gamma^a \Gamma^{bcd}
\label{eq:covmat}
\end{eqnarray}
Low energy anomaly cancellation considerations determine the
finite-dimensional Yang-Mills gauge group to be $O(32)$ \cite{polbook},
$A_{ab}$ is the Neveu-Schwarz
two-form potential which couples to the fundamental string. The 
Ramond-Ramond gauge potentials, distinguished by 
dilaton-independent kinetic terms in $\cal S$, are denoted 
$C_{a_1 \cdots a_p}$, with $p$ even, 
and $0$$\le$ $p$$\le$$10$. They couple to Dpbranes \cite{dbrane}.  
Notice that $C_{(0)}$ is always non-vanishing corresponding
to the presence of Dinstanton charge, and the consequent 
restriction to IB ground states with a maximum of sixteen 
conserved supercharges. Notice also that under a
$T_9$-duality transformation, we map to the generic I$^{\prime}$
ground state, coupling instead to one or more $p$$\pm$$1$-form 
gauge potentials, with $p$ even. This is the massive IIA string
with nonvanishing cosmological constant, upto sixteen unbroken
supersymmetries, and nine-dimensional nonabelian gauge fields
\cite{flux}. Finally, it is clear that we can map from, respectively,
the IB $O(32)$ and I$^{\prime}$ $E_8$$\times$$E_8$ ground states to
the corresponding heterotic ground states which are their weak-strong
coupling duals, by suitable re-identifications of both operators 
and couplings in the action \cite{wit,polbook}.

\vskip 0.1in
The five-index antisymmetric tensor in Eq.\ (\ref{eq:covmateq}) has been
set to zero in the large $N$ limit. The necessity for 
including such a term at finite $N$ comes from closure of the 
matrix Lorentz algebra: 
a non-vanishing commutator, $[\Omega_{abc} ,
{\bf L}_{de}]$, results
from Lorentz transformation of the
3-index spin-connection. For example, in the 
absence of the $C_{(p)}$ and at finite $N$,
$\Gamma^a D_a$ acts on the generic sixteen-component,
Grassmann-valued, $N$-dimensional $SU(N)$ multiplet as follows:
\begin{equation}
\Gamma^a D_a \Psi = \Gamma^a \left ( \Omega_a + \half \Omega_{abc} \Gamma^{bc}
+ \quart \Omega_{abcde} \Gamma^{bcde} \right ) \Psi 
\label{eq:covfive}
\end{equation}
Upon inclusion of the Ramond-Ramond two-form potential, 
$C_{ab}$, the nontrivial commutator with ${\bf L}_{de}$ implies the
further necessity of including {\em both} 
$C_{[0]}$ and $C_{[4]}$ potentials. 
Thus, in the finite $N$ case, the simultaneous presence of all 
even rank Ramond-Ramond potentials follows simply from Lorentz invariance.
We should note that, in the continuum limit, the zero, two-, and four-form 
Ramond-Ramond potentials are known to couple through mixed Chern-Simons 
terms in the presence of worldvolume nonabelian gauge fields \cite{polbook}. 

\vskip 0.1in
It is helpful to verify explicit closure of the finite $N$ symmetry
algebra $\cal G$. For simplicity, we set all Ramond-Ramond potentials
to zero. Begin with the finite $N$ manifestation of continuum Lorentz 
transformations. We introduce an infinitesimal 
hermitian matrix, ${\rm L}_{ab}$,
antisymmetric under the interchange of tangent space indices
$a$,$b$. Keeping terms upto linear in ${\rm L}_{ab}$, it is easy
to verify that each term in ${\cal S}$ is invariant under
the matrix transformations:
\begin{eqnarray}
\delta \chi^i =&&  \Gamma^{ab} {\rm L}_{ab} \chi^i
, \quad \quad
\delta {\bar{\chi}}^i = -  {\bar{\chi}}^i \Gamma^{ab} {\rm L}_{ab}
\cr
\delta \psi_{a} =&&  \Gamma^{bd} {\rm L}_{bd} \psi_a
    + {\rm L}_a^c \psi_c 
\cr
\delta {\bar{\psi}}_c =&& - {\bar{\psi}}_c \Gamma^{ab} {\rm L}_{ab}
    - {\bar{\psi}}_a {\rm L}^a_c
\cr
\delta \lambda =&& \Gamma^{bd} {\rm L}_{bd} \lambda
, \quad
\delta {\bar{\lambda}} =
- {\bar{\lambda}} \Gamma^{ab} {\rm L}_{ab}
\cr
\delta (\Gamma^a D_a ) =&& [ \Gamma^{ab} {\rm L}_{ab} , \Gamma^c D_c ]
\cr
\delta (\Gamma^{ab} D_b ) =&& [ \Gamma^{de} {\rm L}_{de} , \Gamma^{ab} D_b ]
+ {\rm L}^a_d \Gamma^{db} D_b
\cr
\delta (\Gamma^{abc} D_b ) =&& [ \Gamma^{de} {\rm L}_{de} , \Gamma^{abc} D_b ]
- [ \Gamma^{abd} D_b , {\rm L}_d^c ]
\cr
\quad \delta ( \Omega_{a} \Phi ) =&& [ {\rm L}_a^c , \Omega_c ] \Phi 
- \Omega_c {\rm L}^c_a \Phi ,
\cr
\delta ( \Phi \Omega^a ) =&& - \Phi [ \Omega^c , {\rm L}_c^a ] +
 \Phi {\rm L}^a_c \Omega^c
\label{eq:lorent}
\end{eqnarray}
Likewise, consider a $d_G$-plet of infinitesimal real matrices, 
$\{ \alpha^j \}$, each of which takes diagonal $N$$\times$$N$ form. 
Here, $d_G$ is the dimension of the nonabelian gauge group with 
hermitian generators $\{ \tau_j \}$. We can verify that every term in
$\cal S$ is invariant under the Yang-Mills transformations:
\begin{eqnarray}
\delta \chi =&&  i \tau^j \alpha^j  \chi
, \quad
\delta (g A_a^j \tau^j ) = [ \Omega_a , \tau^j \alpha^j ]
\cr
\delta ( \Omega_{a} \Phi ) =&& i \tau^j \alpha^j \Omega_a \Phi
, \quad \quad \delta ( \Phi {\hat{\Omega}}_{a} ) = - i \tau^j \Phi {\hat{\Omega}}_a
 \alpha^j
\label{eq:gaugena}
\end{eqnarray}
Finally, we must include in $\cal S$ the
crucial two-fermion and four-fermion terms 
required by closure of a supersymmetry algebra \cite{fs,br}. 
Namely, given infinitesimal 16-component Grassmann-valued parameters, 
$\eta_1$, $\eta_2$, each of which transforms as a $N$-vector of 
the unitary group $SU(N)$, we must verify that the commutator
of two matrix supersymmetry transformations can be expressed as the 
sum of (i) an infinitesimal tangent space translation with parameter,
$\xi^{a}$$=$${\bar{\eta_1}} \Gamma^{a}
\eta_2$, (ii) an infinitesimal
local Lorentz transformation with parameter
${\rm L}_{bc}$$=$$\xi^{a} \omega_{abc}$, and (iii)
an infinitesimal local gauge transformation with gauge
parameter $\alpha^i$$=$$-g \xi^{a} A_{a}^i$ \cite{fs}.
We infer the following transformation rules:
\begin{eqnarray}
\delta A^i_{\mu}  &&= \half {\bar{\eta}} \Gamma_{\mu} \chi^i ,
\quad \delta E^a_{\mu} = \half {\bar{\eta}} \Gamma^a \psi_{\mu},
\quad \delta \Phi = {\bar{\eta}} \lambda
\cr
\delta \chi &&= - \half  ( \Gamma^{ab} F_{ab} ) \eta
+ \half \left ( {\bar{\eta}} \chi - {\bar{\chi}} \eta \right ) \lambda
- \half ({\bar{\chi}} \Gamma^a \eta ) \Gamma_a \lambda
\cr
\delta \psi_{\mu} &&= D_{\mu} \eta + \half \left (
{\bar{\eta}} \psi_{\mu} - {\bar{\psi}}_{\mu} \eta \right ) \lambda
 - \half  ( {\bar{\psi}}_{\mu} \Gamma^a \eta ) \Gamma_a \lambda
\cr 
\quad && \quad\quad
+ {{1}\over{g^2}} ({\bar{\chi}} \Gamma^{abc} \chi ) \Gamma_{abc} \Gamma_{\mu} \eta
\cr
\delta \lambda &&=  -  \quart (\Gamma^a D_a \Phi  ) \eta +
( H_{abc} - {\bar{\lambda}} \Gamma_{abc} \lambda
\cr
\quad && \quad\quad 
+ {{1}\over{g^2}}  {\rm tr} ( {\bar{\chi}} \Gamma_{abc} \chi )  )
 \Gamma^{abc} \eta
\label{eq:susyt}
\end{eqnarray}
The resulting two-fermi and four-fermi terms that must be included 
in $\cal S$ take the manifestly Lorentz and Yang-Mills invariant form:
\begin{eqnarray}
 {\cal S}_{\rm 2} &&=
2 {\bar{\psi}}_{a} \Gamma^{a} \psi_{b} ( \Omega^{b} \Phi )
- 4 {\bar {\psi}}_{a} \Gamma^{b} \Gamma^{a} \lambda ( \Omega_{b} \Phi)
\cr
&&\quad - \eigt H^{def} [
{\bar \psi}_{a} \Gamma^{[a} \Gamma_{def }\Gamma^{b]} \psi_{b}
+ 4 {\bar \psi}_{a} \Gamma^{a}_{def } \lambda
- 4  {\bar\lambda} \Gamma_{def } \lambda \cr
&&\quad
+ g^2 e^{\Phi} {\bar \chi} \Gamma_{def} \chi ]  
+ \quart g^2 e^{\Phi}  {\bar \chi}^i \Gamma^{d} \Gamma^{ab}
(\psi_{d} +  \thre \Gamma_{d} \lambda ) F_{ab}^i
, \cr
{\cal S}_{\rm 4} &&=
\ninsix {\bar\psi}^{f} \Gamma^{abc} \psi_{f}
 (  {\bar\psi}_{d} \Gamma^{d} \Gamma_{abc} \Gamma^{e} \psi_{e}
      + 2  {\bar{\psi}}^{d} \Gamma_{abc} \psi_{d} \cr
\quad &&\quad - 4  {\bar{\lambda}} \Gamma_{abc} \lambda
          - 4 {\bar{\lambda}} \Gamma_{abc} \Gamma^{d}\psi_{d}
                    ) \cr
&&\quad + \ninsix g^2 e^{\Phi}  ( {\bar{\chi}} \Gamma^{abc} \chi )  (
{\bar{\psi}}_{d} ( 4 \Gamma_{abc} \Gamma^{d}  
\cr
\quad &&\quad + 3  \Gamma^{d} \Gamma_{abc} ) \lambda
   - 2 {\bar{\lambda}}\Gamma_{abc}\lambda - 3\cdot 2^3 ~ H_{abc} ) 
\label{eq:24fermimat}
\end{eqnarray}

\vskip 0.1in
Our final expression for ${\cal S}$ can be simplified 
by introducing $SU(N)$ vectors, $\Psi$, ${\bar{\Psi}} $, 
where 
${\bar{\Psi}} $$\equiv $$({\bar{\lambda}} , {\bar{\psi}}_{a} , {\bar{\chi}}^i )$,
$\Psi $$\equiv$$ (\lambda , \psi_{b} , \chi^j )$.
We will assemble the kinetic and two-fermi terms of ${\cal S}$ in a
$(11$$+$${\rm d}_{\rm G})N $$\times $$ (11$$+$${\rm d}_{\rm G})N $
matrix array denoted $\cal D$.
The four-fermi terms are likewise reassembled
by introducing the matrix arrays $\cal U$, $\cal V$, of size
$(11+{\rm d_G})N$$\times$$(11+{\rm d_G})N$, defined by
Eq.\ (\ref{eq:24fermimat}). This give the remarkably compact result:
\begin{eqnarray}
{\cal S} &&= \half {\bar\Psi} {\cal D} \Psi +
  \quart ({\bar{\Psi}}{\cal U} \Psi )( {\bar{\Psi}} {\cal V} \Psi)
  + \half g^2 e^{\Phi} ~ F^{ab} F_{ab}
\cr
  \quad &&\quad + \half \kap ~ ( {\cal R}
     - \half \Phi \Delta  \Phi
     + 3 e^{2\Phi} ~ H^{abc} H_{abc} )
\label{eq:compact}
\end{eqnarray}
In principle, ${\cal S}$ belongs to a family of 
matrix actions whose members can differ by $O(1/N)$ corrections, but
which share the same infrared fixed point behavior in accordance 
with the principle of Universality Classes \cite{ktohwa}. 

\vskip 0.1in
Our proposed action for matrix theory was motivated largely by symmetry 
considerations. In closing, we recall that quantum $O(1/N)$ 
corrections to the planar action can contribute terms of both quantum and 
nonperturbative origin in the low energy effective field theory. 
In the special case 
of string theories with sixteen supercharges, both will be absent. This 
suggests a useful test of our conjecture by direct computation of the spectrum 
of massive (BPS) string states using the matrix theory formalism. It is
possible that, in these cases, the multiple-scaled variation of the large 
$N$ expansion used in a particular ground state of matrix theory bears 
a simple and direct relationship to the $\alpha^{\prime}$ expansion 
of the low energy string effective action in that same vacuum. It may 
even have an interpretation 
in terms of suitably weighted sums over worldsheets of 
different topology \cite{ktohwa,voic}. 
It will be most interesting to clarify the details of this picture.

\vskip 0.1in
\noindent
{\bf ACKNOWLEDGMENTS} I am grateful to Hikaru Kawai for patient explanations 
of many aspects of planar reduction and the large $N$ limit of matrix theories.
This work was begun during a fruitful visit to Kyoto University, Japan. It 
is funded in part by NSF-PHY-9722394. 
 
\vskip 0.1in \noindent{\bf NOTE ADDED (July 2005):} This is a short 
introduction to the proposal for nonperturbative String/M theory in 
hep-th/0201129. The comments in the {\em Note
Added} to that work apply here as well.

\setcounter{footnote}{0}

\end{document}